\documentclass{trbunofficial}

\usepackage{subcaption}
\usepackage{amsfonts}
\usepackage{amsmath}
\usepackage{ragged2e}
\usepackage{float}
\usepackage[section]{placeins}
\usepackage{bbm}

\begin{document}

\title{Complete Street Design informed by Multi-Agent Simulation}

\TRBauthor{Sharika J. Hegde}{Department of Civil and Environmental Engineering, Northwestern University}{sharikahegde2025@u.northwestern.edu}
\TRBauthor{Amirmohammad Khakpour}{Department of Civil and Environmental Engineering, Northwestern University}{Amirmohammadkhakpour2029@u.northwestern.edu}
\TRBauthor{Pedram Beigi}{Department of Civil and Environmental Engineering, George Washington University}{beigi@email.gwu.edu}
\TRBauthor*{Yu (Marco) Nie}{Department of Civil and Environmental Engineering, Northwestern University}{y-nie@northwestern.edu}
\TRBauthor{Hani S. Mahmassani}{Department of Civil and Environmental Engineering, Northwestern University}{masmah@northwestern.edu}

\AuthorHeaders{Hegde, Khakpour, Beigi, Nie, and Mahmassani}

\maketitle

\section*{Abstract}

\noindent\textbf{Objectives:} Complete Street design is currently guided largely by qualitative principles despite the increasing complexity of multimodal urban traffic. This paper develops a microscopic simulation framework for quantitatively evaluating alternative street designs by explicitly representing heterogeneous road users, including Autonomous Vehicles (AVs), Connected Vehicles (CVs), Human-Driven Vehicles (HDVs), pick-up and drop-off (PUDO) services, and micromobility users. The framework jointly evaluates mobility and safety within a unified testbed.

\noindent\textbf{Methods:} The framework enriches microscopic simulation with  behavioral models for heterogeneous road users. It explicitly captures lateral movements, parking maneuvers, pedestrian crossings and jaywalking, and shared-lane bicycle and scooter interactions. Safety is quantified using a conflict severity measure grounded in collision physics, while candidate street layouts are optimized using a genetic algorithm.

\noindent\textbf{Findings:} Applied to a segment of Belmont Avenue in Chicago, Illinois, the framework shows that dedicated bicycle lanes reduce conflict severity at the expense of vehicular throughput. Increasing AV penetration mitigates congestion by reducing traffic shockwaves while improving interactions with pedestrians relative to CVs and HDVs. The optimization framework efficiently identifies Pareto-efficient street layouts that balance mobility and safety.

\noindent\textbf{Novelty:} The proposed framework integrates heterogeneous behavioral modeling, physics-based conflict assessment, stochastic performance evaluation, and design optimization within a single quantitative framework for Complete Street design.

\noindent\textbf{Practical Applications:} The framework provides transportation planners with a predictive virtual testbed for evaluating street designs and quantifying the tradeoff between mobility and safety before implementation.
\\
\\
\noindent\textit{Keywords}: micro-simulation, complete streets, conflict severity, heterogenous road users

\newpage

\section{Introduction}

Since 1950s urban street design in the U.S. has taken a vehicle-centric approach, prioritizing the speed, capacity, and geometric efficiency for motorists \cite{aashto2018greenbook} at the expense of other road users \cite{dumbaugh2011designing,potts2007lane}.  As technological advances continue to diversify urban street users---from e-scooters and delivery robots to driverless cars---this approach has become increasingly outdated and untenable. 
As a timely response to this pressing challenge, the Complete Streets paradigm promotes design principles that safely and efficiently serve \textit{all} users \cite{house2021complete}.

The challenge of designing for all users has grown more acute in recent years. Growth in pickup-and-drop-off (PUDO) services, bike-share and other micro-mobility modes, and connected and autonomous vehicles (CVs/AVs) has intensified competition for curb space and right-of-way (ROW) \cite{marsden2020parking,reck2021explaining,heineke2019sizing}, introducing new safety concerns and making the case for Complete Streets-oriented design more urgent. Evaluating such designs, however, requires simulation tools capable of representing heterogeneous and dynamic user interactions within constrained urban geometry.

Existing tools fall short of this requirement because the models underlying them were each developed with a particular user type in mind---car-following and lane-changing models for motorized traffic \cite{gipps1981behavioural,kesting2007general}, social-force and related models for pedestrians \cite{helbing1995social}, and analogous behavioral models for cyclists \cite{gavriilidou2019microscopic}. Simulation platforms such as SUMO, VISSIM, Aimsun, and MATSim assemble these mode-specific models and, as a result, must stitch their interactions together through predefined priority rules, fixed crossing zones, or simplified right-of-way assumptions \cite{krajzewicz2012recent,fellendorf2010vissim,barcelo2005dynamic,horni2016matsim}. What gets lost in the stitching is precisely what matters on a real urban street: continuous mutual adaptation, informal negotiation, curbside dynamics, and bicycle-vehicle yielding \cite{jaller2021fighting,oketch2000modeling}. Shared-space models relax strict modal separation and represent a step in the right direction \cite{johora2018modeling,schonauer2012modeling}, but they have not yet matured into a comprehensive evaluation framework. A second, less obvious gap compounds the first: conventional surrogate safety measures were developed for conflicts between vehicles of comparable mass and speed, and extend poorly to interactions among fundamentally different users---pedestrians, cyclists, micro-mobility users, and automobiles---whose large disparities in mass, speed, and vulnerability make proximity-based metrics unreliable indicators of actual safety outcomes.

This paper develops a simulation framework that addresses both gaps.  It represents heterogeneous users negotiating constrained urban space without imposing rigid modal separation, and it quantifies mobility and safety within the same testbed using a conflict severity metric suited to mixed traffic. The framework is demonstrated through two complementary analyses. First, we characterize fundamental relationships among traffic flow, safety, and mobility, and examine how PUDO services, dedicated bike lanes, and AV/CV market penetration affect system performance. The results show that dedicated bike lanes substantially reduce conflict severity relative to shared-lane designs, while even modest levels of autonomous vehicle penetration significantly delay traffic flow breakdown. Second, we apply the framework to a real-world case study in Chicago, Illinois, coupling the multi-agent simulator with a metaheuristic optimization algorithm to identify street layouts that optimally balance safety and mobility according to user-specified weights.

The remainder of this paper is organized as follows. Section 2 presents the simulation methodology and Section 3 discusses the metrics that the simulator produces for evaluating the performance of a complete street design.  In Section 4, we describe the formulation and solution algorithm for the simulation-based complete street design problem. Section 5 reports results from numerical experiments and Section 6 concludes the study with a summary of findings and directions for future research.

\section{Multi-agent simulator}
Our multi-agent simulator models bidirectional, multi-lane, multimodal urban street segments to capture the movements of heterogeneous road users, including AVs, CVs, human-driven vehicles (HDVs), buses, trucks, PUDO vehicles, pedestrians, bicycles, and scooters. By explicitly capturing the complex interactions among these users within a confined urban setting, the simulator provides a quantitative tool for evaluating and designing street layouts that balance mobility and safety in an equitable manner across all users.

The simulator captures a wide range of real-world phenomena, including pedestrian jaywalking, PUDO services obstructing through traffic, and cyclists impeding vehicles when sharing the same lane. It achieves this by integrating conventional vehicular traffic flow models with new behavioral models for non-vehicular users, together with explicit interaction models that govern how different road users influence one another.

We begin by providing an overview of the simulator architecture, including its inputs and outputs, before describing the individual modeling modules in detail.

\subsection{Overview}
The Complete Street micro-simulator takes as its primary input a
bidirectional, multi-lane, multimodal urban street segment through which
heterogeneous road users move. Given a demand scenario and a
supply scenario, the simulator executes the simulation and produces
mobility and safety performance measures.

\subsubsection{Demand and supply scenarios}
Each agent in the simulation is assigned one of three movement types. A
\emph{through movement} takes the agent from one end of the street segment to
the other without stopping. A \emph{crossing movement} applies to
pedestrians, who walk along the sidewalk, cross the street---at a designated
crosswalk or informally---and arrive at a location on the opposite side. A
\emph{PUDO movement} describes a vehicle that enters from one end of the
segment, stops at a predefined curbside location for a specified duration to
perform a pick-up or drop-off, then merges back into traffic and exits at
the other end. The collection of agents, each with an assigned movement type
and schedule, constitutes a \emph{demand scenario}.

A \emph{supply scenario} is defined by the street layout: the number of
lanes of each type, lane widths, sidewalk configuration, crosswalk placement,
and the presence and separation of dedicated bicycle infrastructure.

Together, a demand scenario and a supply scenario fully specify a simulation
run.

\subsubsection{Performance measures}
Given the demand and supply scenarios, the simulator produces, first and
foremost, the trajectories of all agents, from which mobility and safety
metrics are derived. Mobility metrics include flow rate, density, and average
speed, as well as their aggregate relationships---e.g., flow-density and
speed-density curves. Safety metrics quantify the frequency and severity of
safety-critical events, such as incidences of small time-to-collision. A
street design is then evaluated by a weighted average of the mobility and
safety metrics, which serves as the objective function for simulation-driven
design optimization.

\subsection{Behavioral models}

\begin{figure}[htbp]
    \centering
    \includegraphics[width=0.99\textwidth]{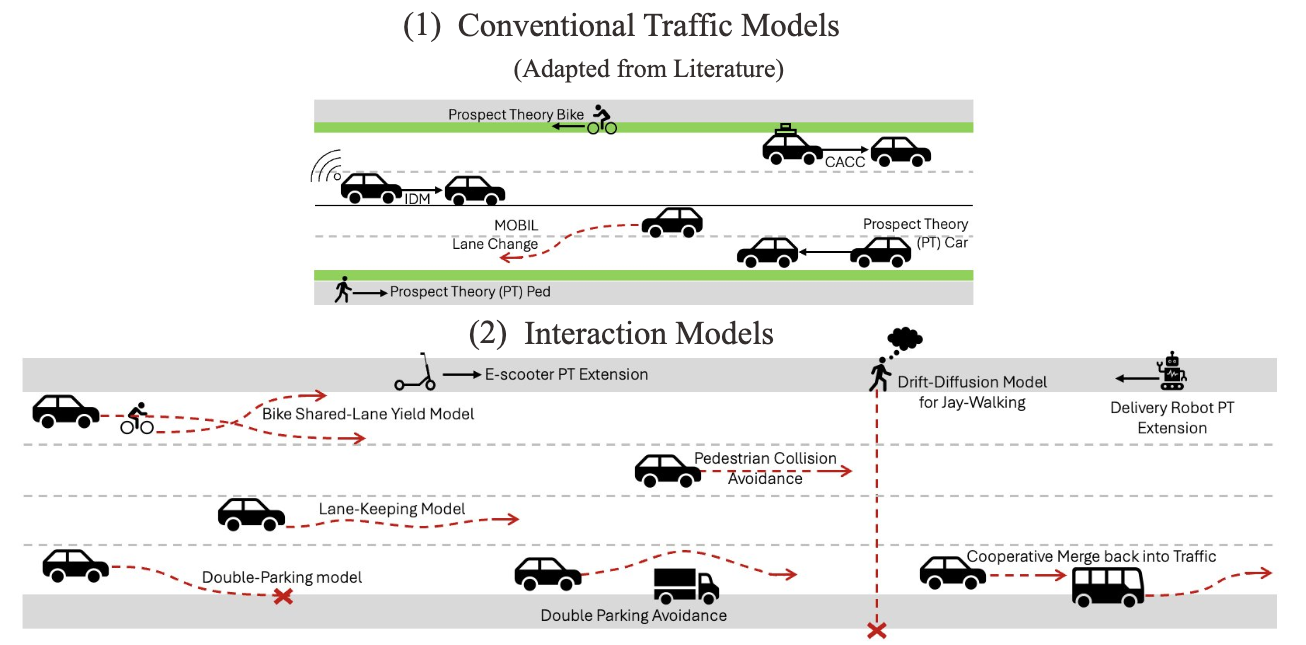}
    \caption{Overview of behavioral models: (1) Conventional traffic models and (2) Interaction models.}
    \label{fig:layout8}
\end{figure}

Behavioral models dictate the operational rules governing agent movements, as
well as interactions both within the same modality and across different agent
types, such as pedestrian-vehicle interactions. As illustrated in
Figure~\ref{fig:layout8}, these models fall into two categories: conventional
traffic flow models and interaction models---most of the latter are, to the
best of our knowledge, new to the literature. Conventional models describe
baseline car-following and lane-changing maneuvers, whereas interaction
models capture complex inter-modal phenomena such as driving around
double-parked vehicles, jaywalking, and bicycle-vehicle negotiation. Both
categories are briefly introduced in the following sections.

\subsubsection{Conventional traffic models}

The simulator adopts conventional microscopic traffic models from the
literature to represent baseline movement behavior; given their wide use, we
do not elaborate on them here. Longitudinal control follows the Cooperative
Adaptive Cruise Control (CACC) model for AVs \cite{vanarem2006impact,
talebpour2016influence}, the Intelligent Driver Model (IDM) for CVs
\cite{treiber2000congested,kesting2010enhanced}, and a prospect-theory-based
car-following model for HDVs, buses, and trucks
\cite{hamdar2008modeling,talebpour2011multiregime}. Lane-changing behavior is
modeled using MOBIL \cite{kesting2007general}. For pedestrians, bicycles, and
scooters, a prospect-theory-based movement extension represents speed and
direction choices under uncertainty
\cite{hamdar2008modeling,li2011analyze,liang2012psychological,
vasic2012cellular}.

\subsubsection{Interaction models}

Interaction models implemented in the current simulator include lane keeping
(cross-lane interactions), double-parking avoidance and cooperative merging
(PUDO activities), pedestrian collision avoidance and drift-diffusion-based
crossing (jaywalking), and bicycle yielding in shared lanes.

Before presenting these models, we define the notations used throughout
this section. The index $i$ denotes the ego agent, $t$ denotes simulation
time, and $\Delta t$ the simulation time step. The longitudinal and lateral
positions of agent $i$ are denoted by $x_i(t)$ and $y_i(t)$, respectively; the
longitudinal speed by $v_i(t)$, the lateral velocity by $v_{y,i}(t)$, and the
acceleration by $a_i(t)$. Agent width is denoted by $w_i$, and distances are
denoted by $d$, with subscripts indicating the specific context, e.g.,
$d_{\text{stop}}$ or $d_{\text{conf}}$. Greek letters denote model-specific
parameters or thresholds, such as damping coefficients, sensitivity
parameters, and safety margins, and are defined when first introduced.

\paragraph{\textit{Lane keeping}. }
The lane-keeping model represents continuous lateral adjustment within a travel lane. The lateral velocity of agent $i$, $v_{y,i}(t+\Delta t)$, is updated based on damping, lane-centering behavior, lateral influence from neighboring vehicles, and stochastic variation:

\[
v_{y,i}(t+\Delta t)=\gamma v_{y,i}(t)+\left[a^{\text{center}}_i(t)+a^{\text{neighbor}}_i(t)+\xi_i(t)\right]\Delta t .
\]
Here, $\gamma$ is the lateral damping coefficient, $a^{\text{center}}_i(t)$ is the lane-centering acceleration, $a^{\text{neighbor}}_i(t)$ captures the lateral influence of nearby vehicles, and $\xi_i(t)$ represents random lateral fluctuation. The lane-centering acceleration is defined as $a^{\text{center}}_i(t)=-\beta_{\text{center}}\left[y_i(t)-y_{\text{center}}(x_i(t))\right]$, where $\beta_{\text{center}}$ controls the strength of the centering force and $y_{\text{center}}(x_i(t))$ denotes the lane centerline at the longitudinal position of the agent.

\paragraph{\textit{Double Parking}.}
The double-parking model represents temporary stopping behavior by PUDO or
delivery vehicles, together with the responses of surrounding traffic to the
resulting obstruction. When a vehicle prepares to stop, the required
stopping distance is computed as $d_{\text{stop}}=v_i^2(t)/(2a_{\max})$,
where $a_{\max}$ is the maximum comfortable deceleration. During the
pull-over maneuver, position is updated as:
\[
x_i(t+\Delta t)=x_i(t)+v_i(t)\cos(\theta_{\text{park}})\Delta t, \qquad
y_i(t+\Delta t)=y_i(t)+v_i(t)\sin(\theta_{\text{park}})\Delta t,
\]
where $\theta_{\text{park}}$ is the drift angle toward the stopping location.
Once the vehicle reaches the stopping position, it enters a dwell state and
acts as a temporary obstruction to surrounding traffic.

This obstruction, in turn, triggers avoidance and merging behavior in
surrounding vehicles. Vehicles approaching a stopped or double-parked vehicle
first attempt to change lanes. If a lane change is infeasible, the vehicle
evaluates whether it can laterally pass the obstruction within the available
roadway width. Passing is permitted only when
$\left|y_{\text{tar}}-y_{\text{obs}}\right| \geq
(w_i+w_{\text{obs}})/2+s_{\text{lat}}$, where $y_{\text{tar}}$ is the target
passing position, $y_{\text{obs}}$ is the lateral position of the
obstruction, $w_{\text{obs}}$ is the obstruction width, and $s_{\text{lat}}$
is the lateral safety margin.

After completing its dwell time, the stopped vehicle attempts to merge back
into traffic. A merge is considered feasible if an approaching follower can
create an acceptable gap through cooperative braking. The required
deceleration of the follower is:
\[
a_{\text{req}}=-\frac{v_f^2(t)}{2(d-s_{\min})},
\]
where $v_f(t)$ is the follower speed, $d$ is the distance to the merge point,
and $s_{\min}$ is the minimum acceptable spacing. Cooperative merging is
feasible when $a_{\text{req}}\geq a_{\text{coop}}^{\min}$, where
$a_{\text{coop}}^{\min}$ is the acceptable cooperative braking threshold.

\paragraph{\textit{Pedestrian crossing}.}
The pedestrian crossing model captures two connected decisions: whether a
pedestrian, cyclist, or micro-mobility user chooses to cross or wait, and
whether an approaching vehicle can safely yield if a crossing is initiated.

Crossing and waiting decisions are represented through a drift-diffusion
process. Following drift-diffusion models of decision making
\cite{ratcliff2008diffusion,pekkanen2022variable}, the accumulated evidence
for crossing evolves as:
\[
\frac{dA_i(t)}{dt}=-\alpha A_i(t)+s_i(t)+\epsilon_i(t).
\]
Here, $A_i(t)$ is the accumulated evidence for agent $i$, $\alpha$ is a
damping coefficient, $s_i(t)$ is the perceived traffic input, and
$\epsilon_i(t)$ is a stochastic noise term. Crossing begins when
$A_i(t)\geq A_{\text{th}}$, where $A_{\text{th}}$ is the crossing threshold.
For multi-lane streets, the perceived input is defined as
$s_i(t)=\sum_{\ell \in \mathcal{L}}\beta_{\ell}s_{\ell}(t)$, subject to
$\sum_{\ell \in \mathcal{L}}\beta_{\ell}=1$, where $\mathcal{L}$ is the set
of approaching lanes, $s_{\ell}(t)$ is the perceived input from lane $\ell$,
and $\beta_{\ell}$ is the corresponding lane weight.

Once a crossing is initiated, approaching vehicles must determine whether
they can safely yield. The required deceleration to stop before the
predicted conflict point is computed as $a_{\text{ped}}=-v_i^2(t)/
(2d_{\text{conf}})$, where $d_{\text{conf}}$ is the distance from the
vehicle to the predicted conflict point. If the required deceleration is
within the acceptable yielding limit, the vehicle yields. Otherwise,
emergency braking or an alternative conflict-avoidance response is
triggered.

\paragraph{\textit{Bike-vehicle interaction}.}

A bicyclist who travel in a vehicle lane may yield laterally when their presence substantially constrains a following vehicle. The imposed disadvantage on the following vehicle is defined as $D_i=\left|a_i^{\text{with bike}}-a_i^{\text{clear}}\right|$, where $a_i^{\text{with bike}}$ is the following vehicle's acceleration when constrained by the bicycle and $a_i^{\text{clear}}$ is the acceleration the vehicle would have under unobstructed conditions. If $D_i>\delta_{\text{yield}}$, the bicycle yields laterally toward the lane edge, where $\delta_{\text{yield}}$ is the yielding threshold.

During the yielding maneuver, the bicycle position is updated using:
\[
x_i(t+\Delta t)=x_i(t)+v_i(t)\cos(\theta_{\text{yield}})\Delta t, \qquad
y_i(t+\Delta t)=y_i(t)+v_i(t)\sin(\theta_{\text{yield}})\Delta t,
\]
where $\theta_{\text{yield}}$ is the drift angle toward the lane edge. The maneuver consists of three stages: yielding laterally, holding position while the vehicle passes, and returning to the original riding position.

\subsubsection{Decision Hierarchy}

The simulator coordinates behavioral models through a hierarchical decision
framework, executed by each agent at every timestep. The framework
distinguishes two priority levels. At the top, agents first check for
safety-critical conflicts; if one is detected, a conflict-resolution behavior
takes immediate precedence over all other decisions to prevent an imminent
collision---for example, a vehicle sharply decelerating to avoid a crossing
pedestrian, overriding its current car-following target. In the absence of
such a conflict, the agent proceeds through three ordinary decision layers,
from broadest to most immediate. The strategic layer determines whether and
when to initiate a stopping, parking, or crossing maneuver toward the agent's
target destination. The tactical layer governs lane selection, lane-changing,
and yielding decisions conditional on the strategic goal. The operational
layer governs moment-to-moment acceleration and car-following behavior
conditional on the tactical decision.  This decision structure ensures that
routine movement is planned top-down, while safety-critical events can
interrupt this planning at any level.

\section{Performance Measures}
\label{Performance_measures}

The microscopic simulator produces a variety of performance measures for evaluating street designs. In what follows, we first present safety measures, including conflict frequency and severity, followed by operational measures such as throughput and traffic density. 

\subsection{Conflict frequency and severity measures}

Conflicts are identified using the generalized time-to-collision (GTTC), following standard surrogate safety analysis practice \cite{hayward1972near,gettman2003surrogate,gettman2008ssam}. While GTTC is effective for detecting potential conflicts, it only measures the remaining time to collision assuming that all road users maintain their current trajectories and therefore provides no indication of the consequences of a collision. To complement GTTC, we introduce a separate severity measure based on the estimated collision energy. This metric captures the asymmetric risks associated with different conflict types, allowing, for example, a pedestrian--vehicle conflict to be distinguished from a vehicle--vehicle conflict even when they exhibit similar GTTC values. The proposed conflict detection and severity measures are described below.

\subsubsection{Conflict identification using GTTC}

At each timestep, all agent pairs on the same link are evaluated. The generalized time to collision is calculated by:
\[GTTC_{ij}
=
\frac{d_{ij}^{\text{adjusted}}}{v_{ij}^{\text{approach}}},\]
where \(d_{ij}^{\text{adjusted}}\) is the size-adjusted separation distance and \(v_{ij}^{\text{approach}}\) is the relative closing speed. A conflict is identified when $0 < GTTC_{ij} \leq T_{\text{crit}}$, with $T_{\text{crit}} = 3~\text{s}$. The conflict is tracked until the agents are no longer converging or the GTTC exceeds the threshold.

\subsubsection{Severity metric based on impact energy}

To define the severity of a conflict based on impact energy, consider two agents \(i\) and \(j\) and define the reduced mass is defined as $\mu_{ij}
=
\frac{m_i m_j}{m_i + m_j}$, where \(m_i\) and \(m_j\) are the masses for the agents. The impact energy is then calculated as:
\[E_{ij}
=
\frac{1}{2}
\mu_{ij}
v_{ij,\text{rel}}^2,
\]
where \(v_{ij,\text{rel}} = \lVert \mathbf{v}_i - \mathbf{v}_j \rVert\). Energy is allocated to agent $i$ as $E_i
=
E_{ij}
\frac{m_j}{m_i+m_j},
$ and to agent $j$ as $
E_j
=
E_{ij}
\frac{m_i}{m_i+m_j}.$ The normalized severity for agent \(i\) can then be calculated as $S_i
=
\frac{E_i}{E_{\text{threshold},i}}$, where \(E_{\text{threshold},i}\) is the injury-risk threshold for that agent type \cite{lubbe2022safe}. Finally, the conflict severity for this pair of conflicting agents is defined as: $S_{ij}(t)
=
\max
\left(
S_i(t), S_j(t)
\right),$ which prioritizes the more severely affected user and is consistent with safety-oriented Complete Street design \cite{johansson2009vision}.

For each conflict episode \(c\) in the simulation, we record its severity as the maximum value achieved within the duration of the episode $\mathcal{T}_c$ , i.e., $S_c
=
\max_{t \in \mathcal{T}_c}
S_{ij}(t)$. The aggregate simulation-level severity, one of the key performance measures, is calculated as $S_{\text{total}}
=
\sum_{c \in \mathcal{C}}
S_c$, where \(\mathcal{C}\) is the set of all detected conflicts during the simulation horizon.

\subsection{Operational Measures}

The  operational measures produced by the simulator include, but are not limited to: throughput, density, and average speed of all vehicles. In this paper, throughput is chosen as the variable used to analyze the operational efficiency of a specific street layout. Throughput is defined as the total number of vehicles exiting the simulation by the end of the simulation, i.e.,
\[T_r(s)
=
\sum_{i \in \mathcal{A}_r}
\mathbbm{1}
\left(
x_i(t_{\mathrm{end}}) \geq L
\right),\] where \(\mathcal{A}_r\) is the set of all types of vehicles in replication \(r\), \(x_i(t_{\mathrm{end}})\) is the final position, \(L\) is the corridor length, and \(\mathbbm{1}(\cdot)\) is an exit indicator. 

\section{Design Problem}
\label{Design_Problem}
The street design problem is formulated as a simulation-based multi-objective
optimization problem. Let $s\in\Omega$ denote a feasible street design, where
$\Omega$ is the set of layouts satisfying geometric, right-of-way, and
traffic-design constraints. For each candidate design, the 
microscopic simulator described in the previous sections estimates its
expected operational and safety performance for a given demand profile.

\begin{figure}[htbp]
    \centering
    \includegraphics[width=0.85\textwidth]{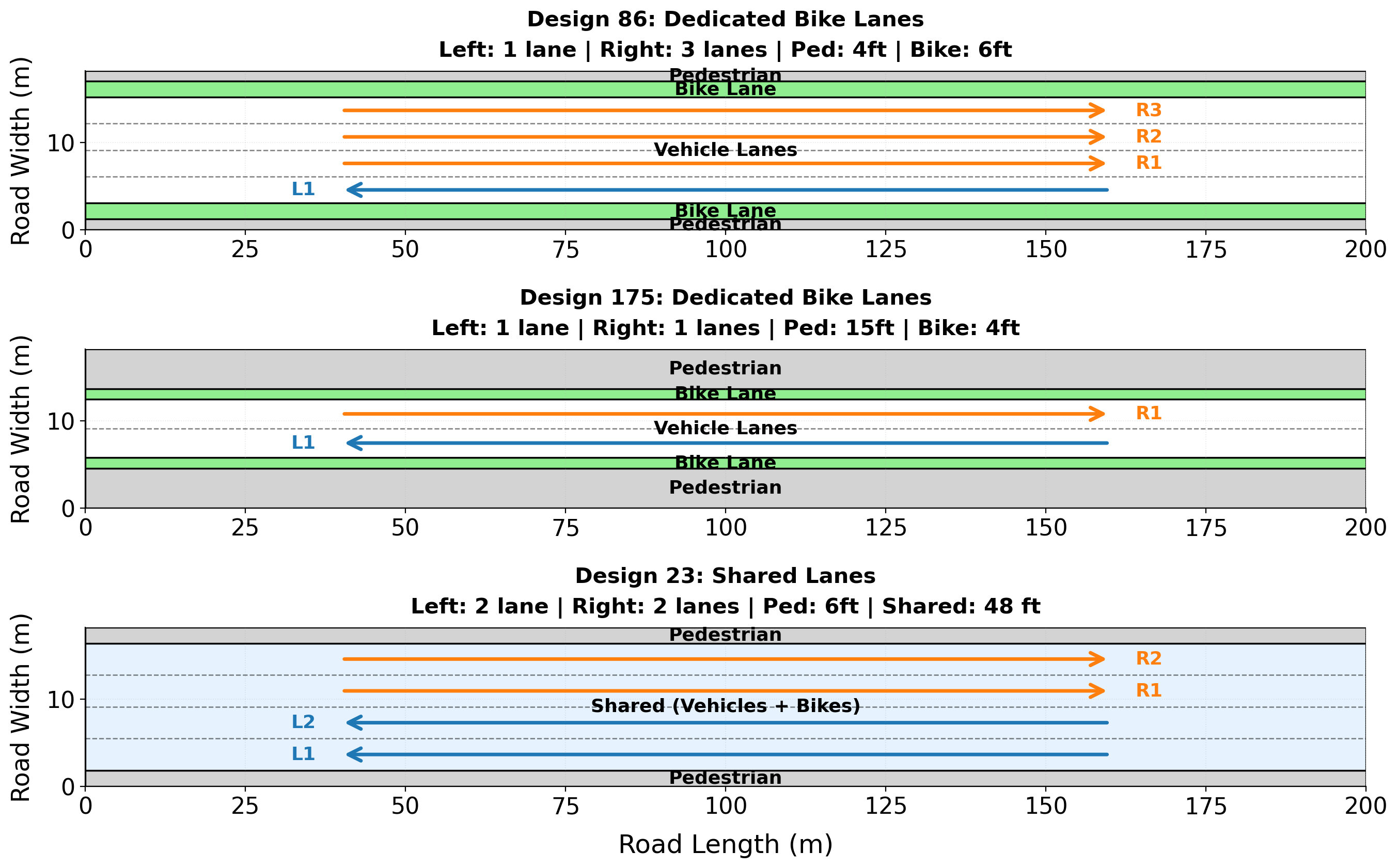}
    \caption{Sample candidate street designs.}
    \label{fig:layout6}
\end{figure}

A candidate design is characterized by the allocation of right-of-way among
vehicle lanes, bicycle facilities, and sidewalks. In this study, the design
variables are
\[
s
=
\left(
N_E,
N_W,
w_{\mathrm{veh}},
w_{\mathrm{bike}},
w_{\mathrm{sidewalk}},
z_{\mathrm{bike}}
\right),
\]
where $N_E$ and $N_W$ denote the numbers of eastbound and westbound vehicle
lanes, $w_{\mathrm{veh}}$ is the vehicle lane width,
$w_{\mathrm{bike}}$ is the bicycle-facility width,
$w_{\mathrm{sidewalk}}$ is the sidewalk width, and
$z_{\mathrm{bike}}\in\{0,1\}$ indicates whether a dedicated bicycle facility
is provided. Figure~\ref{fig:layout6} illustrates several representative
layouts from the feasible design space.

The design seeks to simultaneously maximize expected throughput and minimize
expected conflict severity. Formally, the problem can be expressed as
\[
\begin{aligned}
\max_{s\in\Omega}
\quad
&
\left(
\bar T(s),
-\bar S(s)
\right),
\end{aligned}
\]
where $\bar T(s)$ and $\bar S(s)$ denote the expected throughput and expected
conflict severity obtained from the stochastic simulation model. Because
these objectives generally conflict, the solution is characterized by a
Pareto frontier rather than a single optimum.

To generate Pareto-efficient solutions, we adopt a weighted-sum
scalarization,
\[
J_{\alpha}(s)
=
\alpha
\frac{\bar T(s)}{T_{\max}}
-
(1-\alpha)
\frac{\bar S(s)}{S_{\max}},
\]
where $T_{\max}$ and $S_{\max}$ are normalization constants, and
$\alpha\in[0,1]$ specifies the relative emphasis placed on operational
efficiency and safety. For a given value of $\alpha$, the optimization problem
becomes
\[
s^*
=
\arg\max_{s\in\Omega}
J_{\alpha}(s).
\]
All candidate layouts must satisfy the available right-of-way constraint,
\[
N_Ew_{\mathrm{veh}}
+
N_Ww_{\mathrm{veh}}
+
2w_{\mathrm{sidewalk}}
+
2z_{\mathrm{bike}}w_{\mathrm{bike}}
\le
W_{\mathrm{ROW}},
\]
where $W_{\mathrm{ROW}}$ is the available right-of-way width. Additional
geometric and traffic-design constraints further define the feasible design
space. 

\subsection{Scenario generation and performance evaluation}

The performance of a street design cannot be reliably assessed from a single
simulation run because urban traffic is inherently stochastic. Even under the
same geometric design and average demand, vehicle arrivals, travel speeds,
pedestrian crossing behavior, curbside activities, and the resulting
interactions vary from one realization to another. Consequently, each
candidate street layout is evaluated over multiple scenarios, and
its performance is characterized by the expected values of the resulting
performance measures rather than by any individual simulation outcome.

A baseline multimodal demand  for a multi-modal street corridor ---  traffic volume, pedestrian, curbside activities, pedestrian crossings --- may be obtained from field observations.  These observations are converted into three
demand schedules describing (i) through movements, including entry rates and
desired speeds of continuously moving users; (ii) crossing activities,
including the timing and locations of pedestrian and micromobility crossings
and jaywalking events; and (iii) curbside activities, including the timing,
location, and dwell duration of stopping vehicles.

To generate stochastic traffic scenarios, each demand schedule is perturbed
independently across Monte Carlo replications. Through-moving agents vary in
arrival time, desired speed, and realized demand level; stopping vehicles vary
in stopping time, stopping location, and dwell duration; and crossing events
vary in both timing and location. These perturbations capture the natural
variability of urban street activity while preserving the observed average
traffic conditions.

For each candidate street layout $s$, $R$ independent simulation replications
are performed. Let $T_r(s)$ and $S_r(s)$ denote the throughput and conflict
severity obtained from replication $r$, respectively. The expected
performance of the design is estimated by

\[
\bar T(s)
=
\frac{1}{R}
\sum_{r=1}^{R}
T_r(s),\;
\mbox{and}\;
\bar S(s)
=
\frac{1}{R}
\sum_{r=1}^{R}
S_r(s).
\]
These expected values provide the primary basis for comparing alternative
street designs, while the variability across replications offers an indication
of each design's robustness to uncertain demand and traveler behavior.

\subsection{Solution procedure}

The appropriate solution strategy depends on the size of the feasible design
space. When the number of feasible layouts is sufficiently small, exhaustive
enumeration provides the most transparent solution approach. Every design
$s\in\Omega$ is evaluated, allowing the complete objective-space
representation to be constructed and the Pareto frontier to be identified
directly. Enumeration also provides a useful benchmark because it guarantees
that no feasible design is omitted from the comparison. In the case
study reported in this paper, the approximately 200 feasible layouts are sufficiently few to permit
complete enumeration.

For larger or more detailed design problems, however, the number of feasible
layouts may grow rapidly as additional geometric, operational, or traffic-
control decisions are introduced. Because each objective evaluation requires
multiple stochastic simulation runs, exhaustive enumeration may then become
computationally prohibitive. In such cases, population-based metaheuristics,
such as genetic algorithms (GAs), provide a practical alternative \cite{holland1975adaptation,goldberg1989genetic}. A GA represents each candidate street
layout as a chromosome and evolves a population of layouts through selection,
crossover, mutation, and elitist retention. Candidate fitness is evaluated
using the same Monte Carlo simulation procedure, while successive generations
concentrate the search on promising regions of the design space.

In our case study, enumeration is used to recover the full set of feasible
alternatives and establish the true Pareto frontier. A GA is then applied to
the same problem to examine whether high-performing and Pareto-efficient
designs can be identified with substantially fewer simulation evaluations.
This comparison provides evidence regarding the suitability of the
metaheuristic approach for future applications involving much larger design
spaces.

\section{Numerical results}
\subsection{Traffic flow dynamics}
To validate the multi-modal capabilities of the proposed simulator, we perform a series of numerical experiments at both the microscopic and macroscopic levels. 
The validation is conducted on a simplified bidirectional corridor consisting of one lane in each direction with constant vehicle inflow. Three homogeneous traffic streams are considered separately, consisting entirely of AVs, CVs, or HDVs. To introduce recurring disturbances, five pedestrian crossing events are generated at the same midblock location (\(x=100~\mathrm{m}\)), requiring approaching vehicles to yield before accelerating back to their desired speeds. To isolate the effects of vehicle-following and pedestrian interactions, no curbside stopping, bicycle traffic, or lane-changing behavior is included in this validation experiment.

\begin{figure}[htbp]
    \centering
    \begin{subfigure}{0.49\textwidth}
        \centering
        \includegraphics[width=\textwidth]{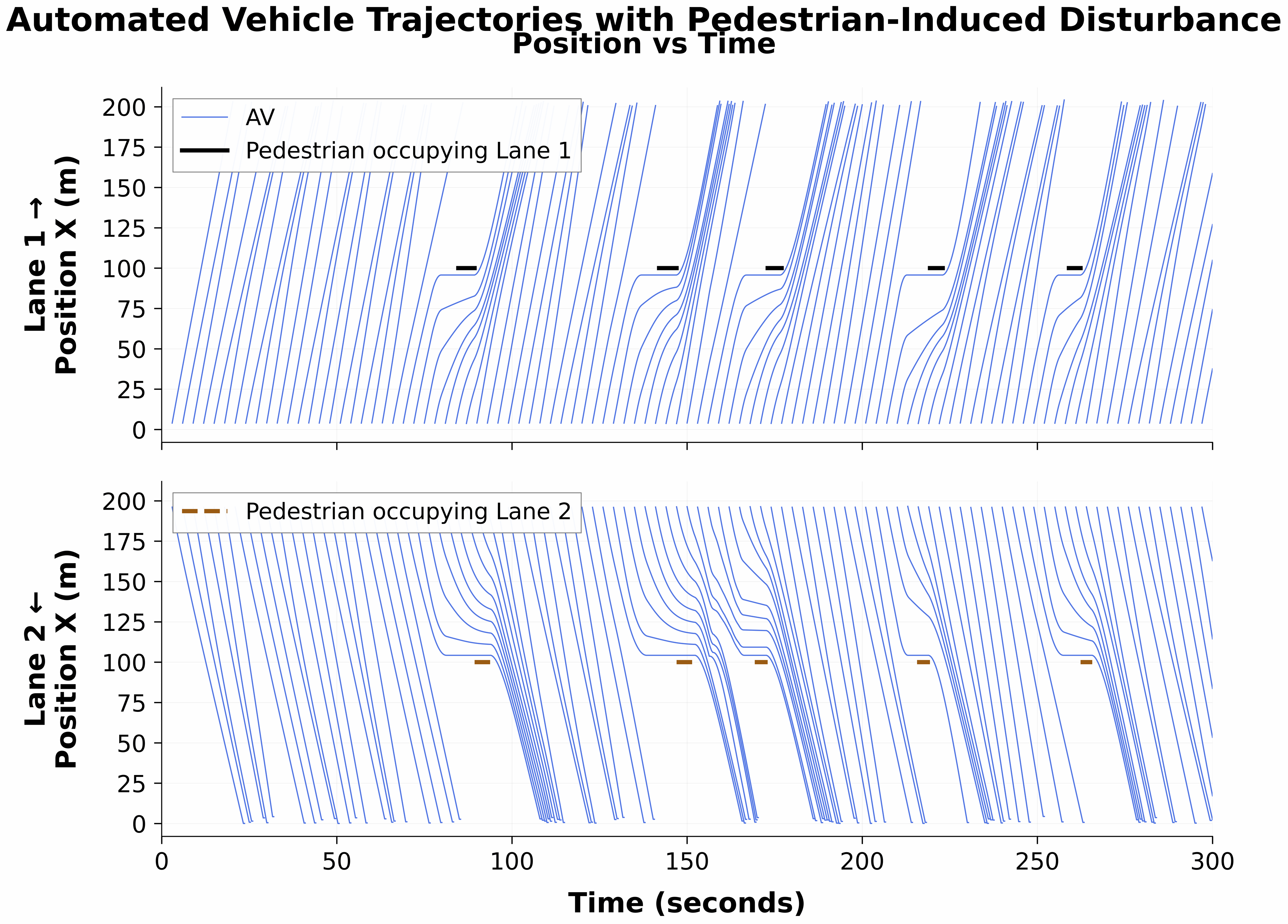}
        \caption{AV stream}
        \label{fig:shockwave_av}
    \end{subfigure}
    \hfill
    \begin{subfigure}{0.49\textwidth}
        \centering
        \includegraphics[width=\textwidth]{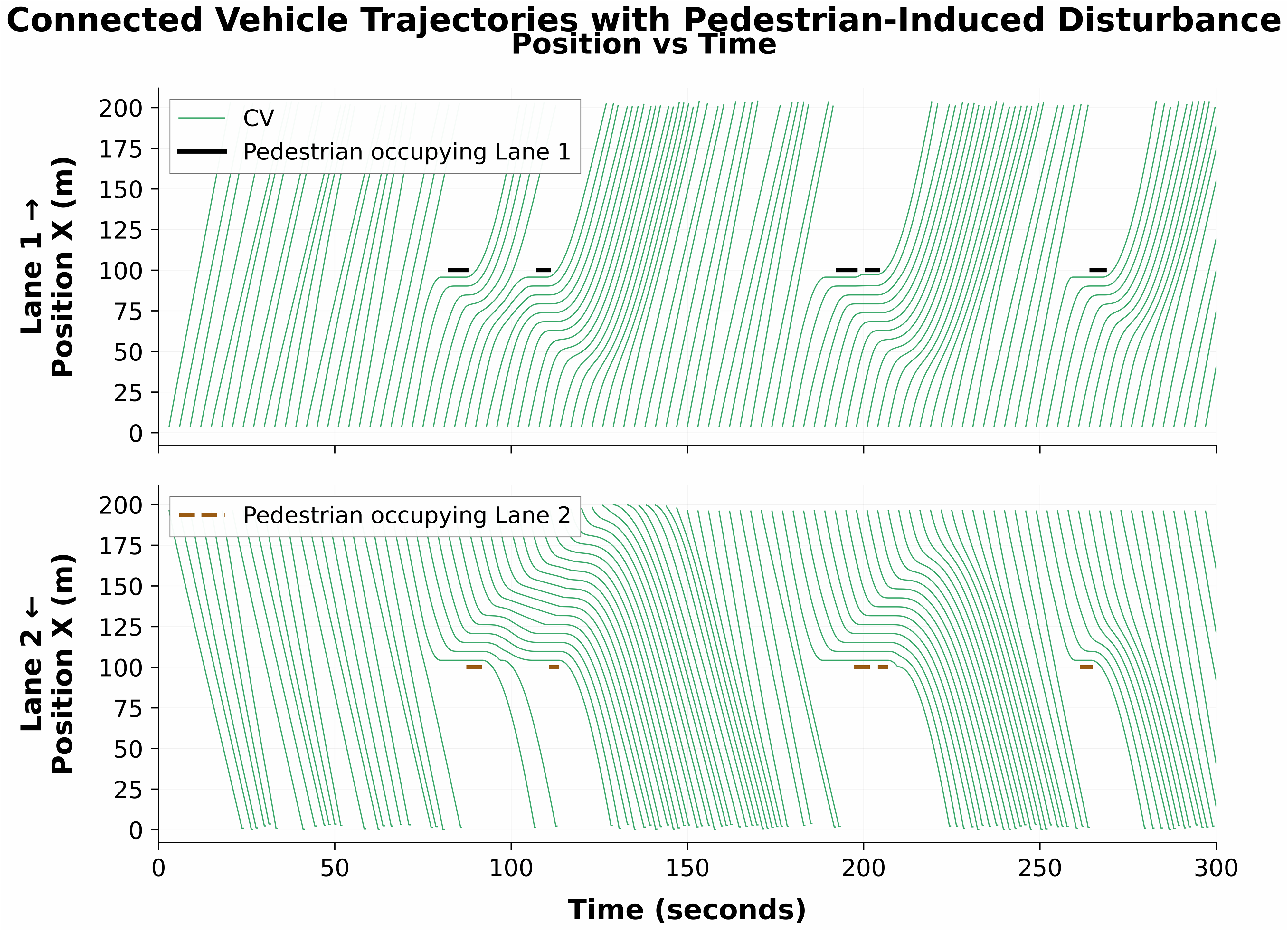}
        \caption{CV stream}
        \label{fig:shockwave_cv}
    \end{subfigure}
    
    \vspace{0.5em}
    
    \begin{subfigure}{0.49\textwidth}
        \centering
        \includegraphics[width=\textwidth]{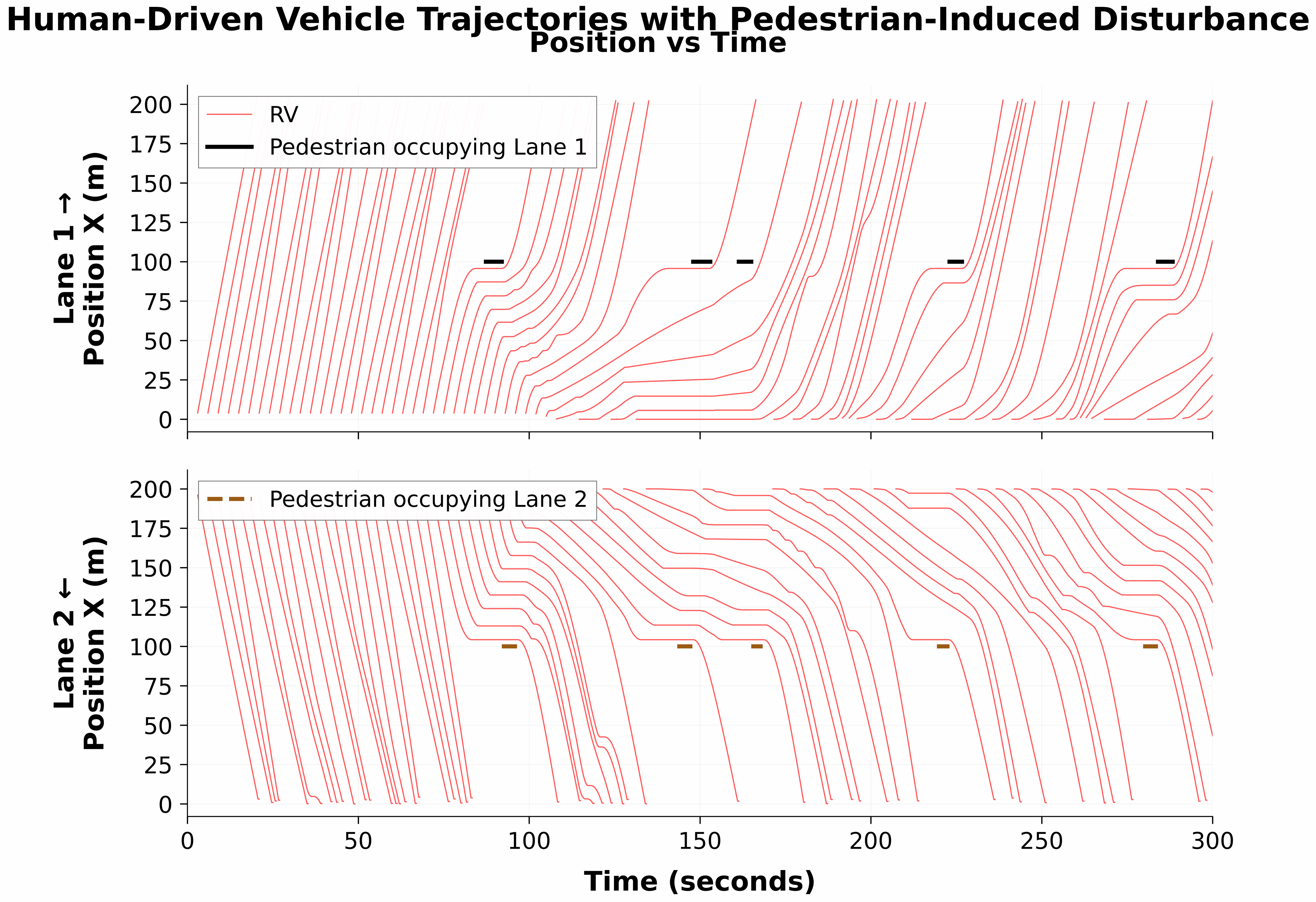}
        \caption{HDV stream}
        \label{fig:shockwave_hdv}
    \end{subfigure}
    \hfill
    \begin{subfigure}{0.49\textwidth}
        \centering
        \includegraphics[width=\textwidth]{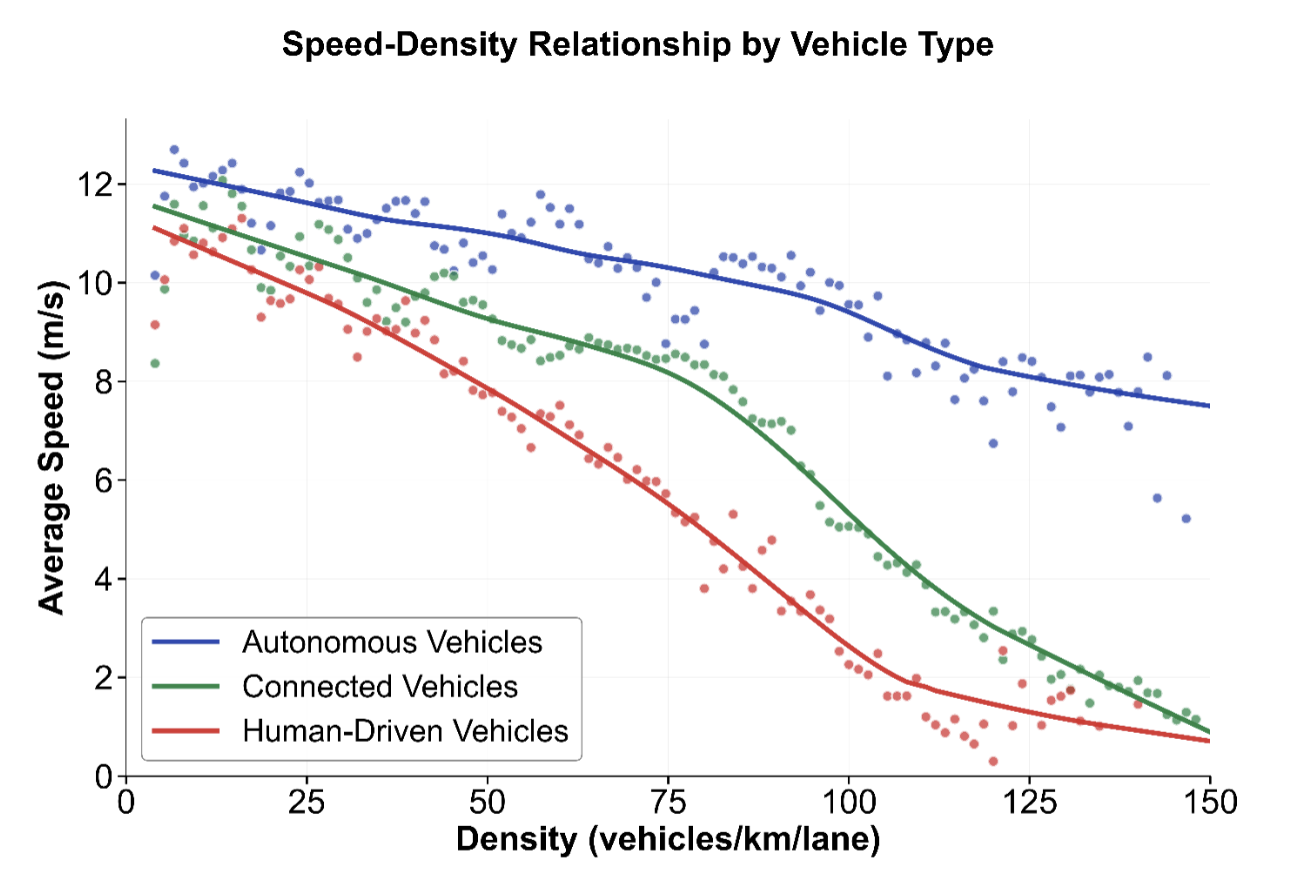}
        \caption{Speed-density relationship}
        \label{fig:speed_density}
    \end{subfigure}
    
    \caption{Trajectory-based shockwave propagation under pedestrian disturbances: (a) AV stream, (b) CV stream, (c) HDV stream, (d) Speed-density relationship for each of these streams not interacting with pedestrians.}
    \label{fig:validation_shockwaves}
\end{figure}

Figure~\ref{fig:shockwave_av}, ~\ref{fig:shockwave_cv}, and ~\ref{fig:shockwave_hdv}  illustrates the simulated vehicle trajectories under the repeated pedestrian crossing events. Each crossing temporarily interrupts the traffic stream, producing localized queues that propagate upstream as backward-moving shockwaves. As pedestrians clear the roadway, the queues gradually dissipate and traffic recovers. The simulator reproduces these well-established traffic phenomena across all three traffic streams while also capturing their distinct behavioral characteristics. AVs exhibit the smoothest recovery following each disturbance, CVs display intermediate behavior, and HDVs produce greater variability because of heterogeneous human driving behavior.

Figure~\ref{fig:speed_density} presents the corresponding speed-density relationships computed from the simulated trajectories. Traffic density and flow are estimated using Edie's generalized definitions \cite{edie1963discussion} over space--time regions. The resulting curves exhibit the expected macroscopic relationship: average speed remains high under light traffic conditions and decreases progressively as density increases. The maximum simulated speed is limited to \(12~\mathrm{m/s}\), consistent with typical urban street operating conditions. Taken together, the trajectory and speed--density results demonstrate that the proposed microscopic interaction model reproduces both realistic local vehicle responses and plausible aggregate traffic-flow behavior, providing confidence in its application to the controlled experiments and street-design evaluations presented in the following sections.

\subsection{Case study}
\subsubsection{Study Area}

Our case study is based  a high-activity, 200-meter segment of Belmont Avenue in Chicago, Illinois, located adjacent to the Belmont CTA Red, Brown, and Purple Line transit station. Characterized by a dense mixture of retail, commercial, and hospitality land uses, this site experiences continuous, competing demands for right-of-way. Specifically, the study area captures a diverse range of multimodal behaviors, including through-vehicle traffic, high-volume pedestrian flows, bicycle and micro-mobility activity, mid-block crossings, and curbside PUDO maneuvers. Consequently, this segment serves as an ideal testbed for evaluating Complete Street designs and analyzing the operational and safety trade-offs inherent in constrained, high-density urban environments. Field data is collected through a site visits conducted during weekday peak periods, capturing vehicles volumes, pedestrian, and cyclist counts, curbside PUDO activity, and midblock crossing locations. These observations were supplemented with aerial imagery and GIS data to establish accurate lane geometry, sidewalk widths, and curb configurations for the segment.  

\subsubsection{Monte Carlo convergence}
\begin{figure}[htbp]
    \centering
    \begin{subfigure}[b]{0.495\textwidth}
        \centering
        \includegraphics[width=\textwidth]{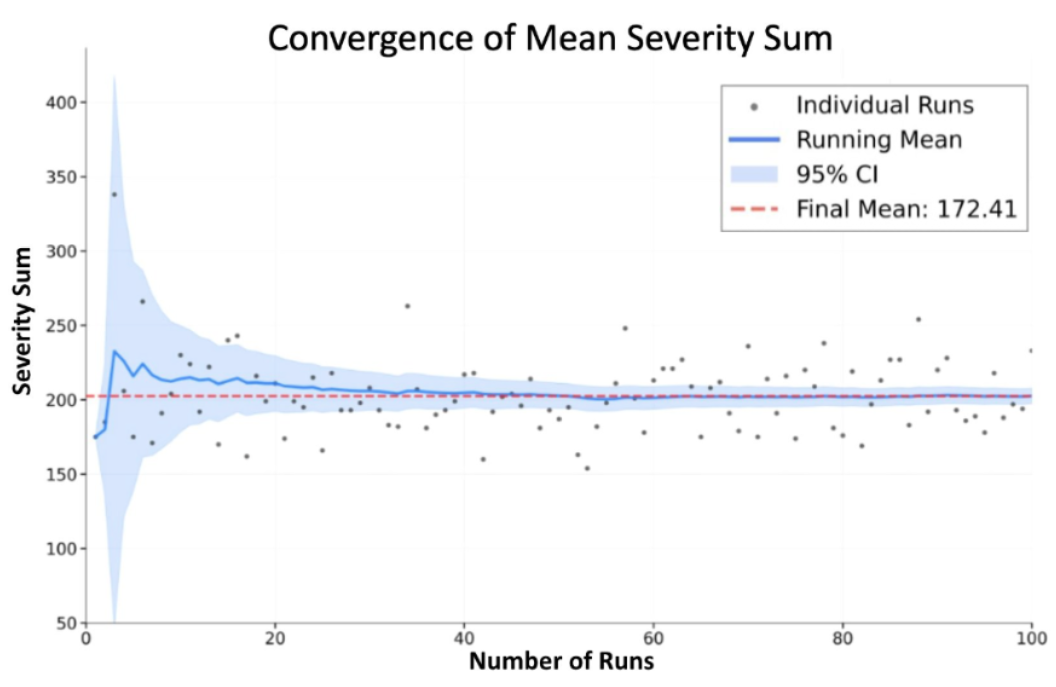}
        \caption{Aggregate conflict severity.}
        \label{fig:layout12}
    \end{subfigure}
    \hfill
    \begin{subfigure}[b]{0.495\textwidth}
        \centering
        \includegraphics[width=\textwidth]{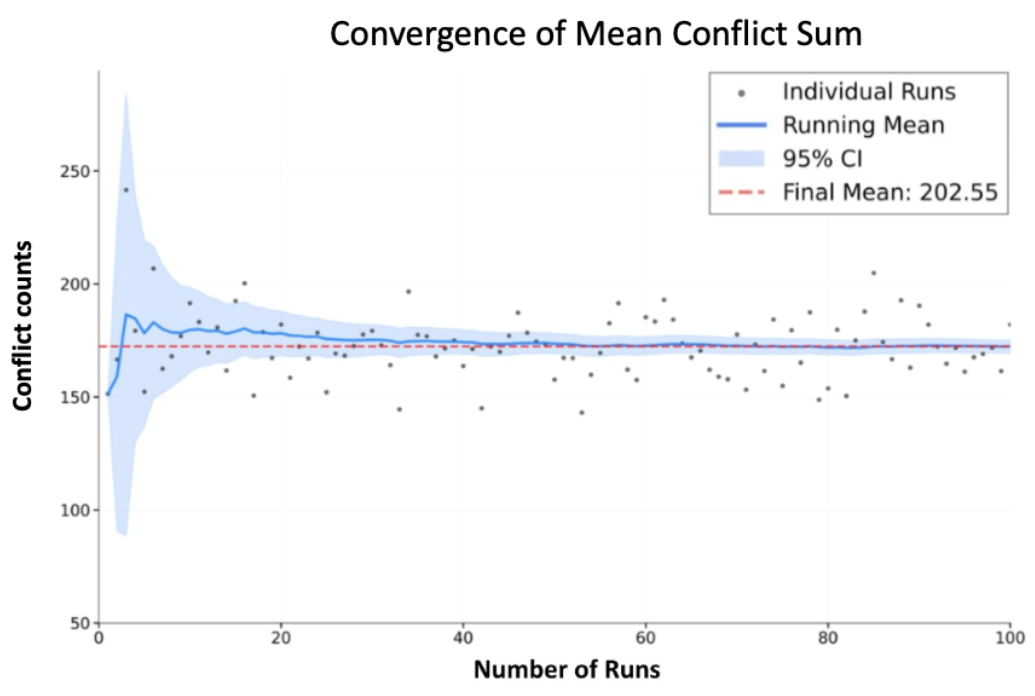}
        \caption{Mean conflict counts.}
        \label{fig:layout13}
    \end{subfigure}
    \caption{Convergence of Monte Carlo performance estimates.}
    \label{fig:layout12_13}
\end{figure}

The first experiment examines the convergence of the Monte Carlo performance estimates. Figures~\ref{fig:layout12} and~\ref{fig:layout13} show the aggregate conflict severity and the running mean of conflict counts as the number of simulation replications increases. Both measures stabilize after approximately 40 replications, indicating that the estimated performance metrics have converged for the case study corridor.

This result justifies the use of repeated stochastic simulation for performance evaluation. Because each replication involves different realizations of street demands, individual simulation outcomes inevitably vary. Averaging over a sufficient number of replications produces stable estimates of the expected performance, thereby enabling reliable comparisons among alternative street designs.

\subsubsection{Performance of candidate street designs}

\begin{figure}[htbp]
    \centering
    \includegraphics[width=0.95\textwidth]{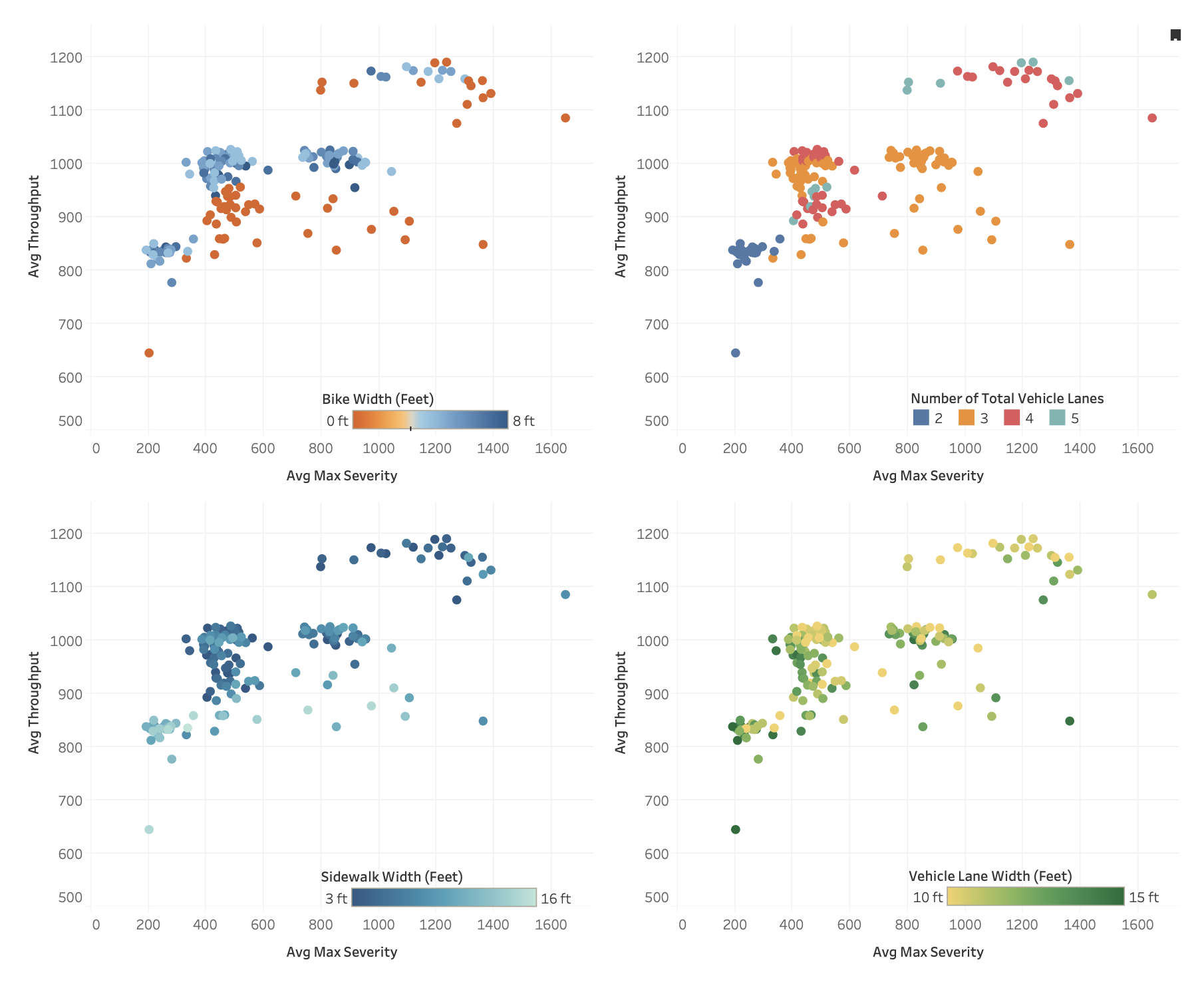}
    \caption{Performance measures of candidate street designs, with conflict severity on the x-axis and throughput on the y-axis. Colors in each panel correspond to the value of a design feature.}
    \label{fig:design_space}
\end{figure}

Figure~\ref{fig:design_space} shows the objective-space representation of the feasible street layouts, with aggregate conflict severity on one axis and throughput on the other. Each point represents one feasible design alternative evaluated using Monte Carlo simulation. The distribution of points demonstrates that design variables affect safety and throughput differently. Some layouts provide high throughput but also produce higher severity, while others reduce severity at the cost of lower throughput.

The scatter plots show the trade-off between average throughput and average conflict severity, where designs located toward the upper left provide the most desirable balance of high throughput and low severity. The clearest patterns are associated with bike-lane width and the number of vehicle lanes. Designs with dedicated bike lanes, particularly moderate-width bike lanes, appear more frequently in the lower-severity region while maintaining reasonable throughput, suggesting that separating cyclists from vehicle traffic can improve safety without substantially reducing operational performance. Lane count creates a stronger trade-off: designs with more vehicle lanes generally achieve higher throughput but also experience greater conflict severity, while two- and three-lane designs tend to provide safer and more balanced alternatives.

Sidewalk width and vehicle-lane width have weaker and less consistent effects. Wider sidewalks are often associated with lower throughput, likely because they use space that could otherwise be allocated to vehicle or bike facilities, but they do not produce a clear additional reduction in the conflict-severity measure. Similarly, wider vehicle lanes appear somewhat more frequently in the lower-severity region, while narrower lanes are more common among high-throughput designs, although considerable overlap exists.

The results show that Complete Street performance cannot be inferred from a single geometric feature alone. Lane allocation, lane width, sidewalk width, and bicycle facility type interact with demand and behavior to produce system-level outcomes. Dedicated bicycle facilities tend to reduce severe bicycle--vehicle interactions by separating incompatible movements, while additional vehicle capacity can improve throughput but may also introduce more lane-changing and passing interactions. These patterns illustrate why simulation-based evaluation is useful for understanding the combined effects of geometry, demand, and heterogeneous agent behavior.

\begin{figure}[htbp]
    \centering
    \includegraphics[width=0.6\textwidth]{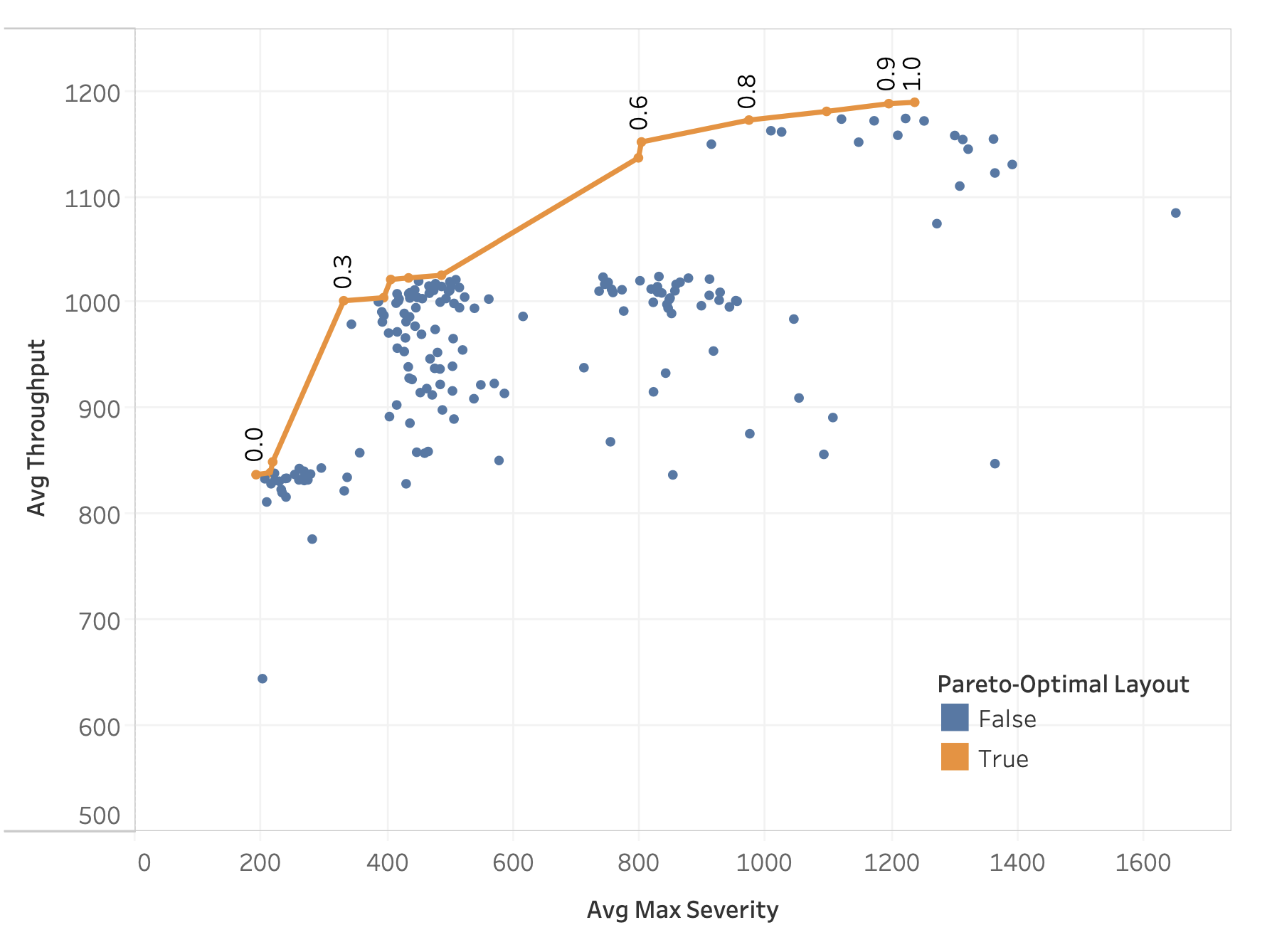}
    \caption{Pareto-optimal frontier illustrating the trade-off between vehicle 
    throughput and conflict severity. Each point corresponds to a non-dominated 
    candidate layout. The parameter $\alpha \in [0,1]$ on Pareto-optimal frontier indicates each layout's position along the safety--throughput trade-off, with $\alpha \to 0$ denoting safety-prioritized designs and $\alpha \to 1$ denoting throughput-prioritized designs.}
    \label{fig:pareto_frontier}
\end{figure}

Figure~\ref{fig:pareto_frontier} shows the Pareto frontier of the candidate layouts. Each point on the frontier represents a Pareto-optimal layout: a layout is Pareto-optimal if no other feasible layout achieves throughput and severity that are both at least as good, with at least one objective strictly better. The frontier therefore highlights the trade-off between safety and operational efficiency. The number of each point on the Pareto-frontier is $\alpha \in [0,1]$ defined in Section ~\ref{Design_Problem}, with $\alpha$ near 0 indicating a safety-prioritized layout (low severity, lower throughput) and $\alpha$ near 1 indicating a throughput-prioritized layout (high throughput, higher severity). Intermediate values of $\alpha$ correspond to compromise layouts that balance both objectives.

%

The Pareto frontier is useful because it avoids treating street design as a single-objective problem. Instead of identifying one universally optimal layout, the framework identifies a family of high-performing alternatives. A planner can then select among these alternatives depending on the desired balance between safety and throughput. The weighting parameter \(\alpha\) in the scalar objective provides one way to trace this frontier and identify designs consistent with local priorities.

\subsubsection{Performance of genetic algorithm}

\begin{figure}[htbp]
    \centering
    \centering
\includegraphics[
    width=0.65\textwidth,
    height=0.7\textheight,
    keepaspectratio
]{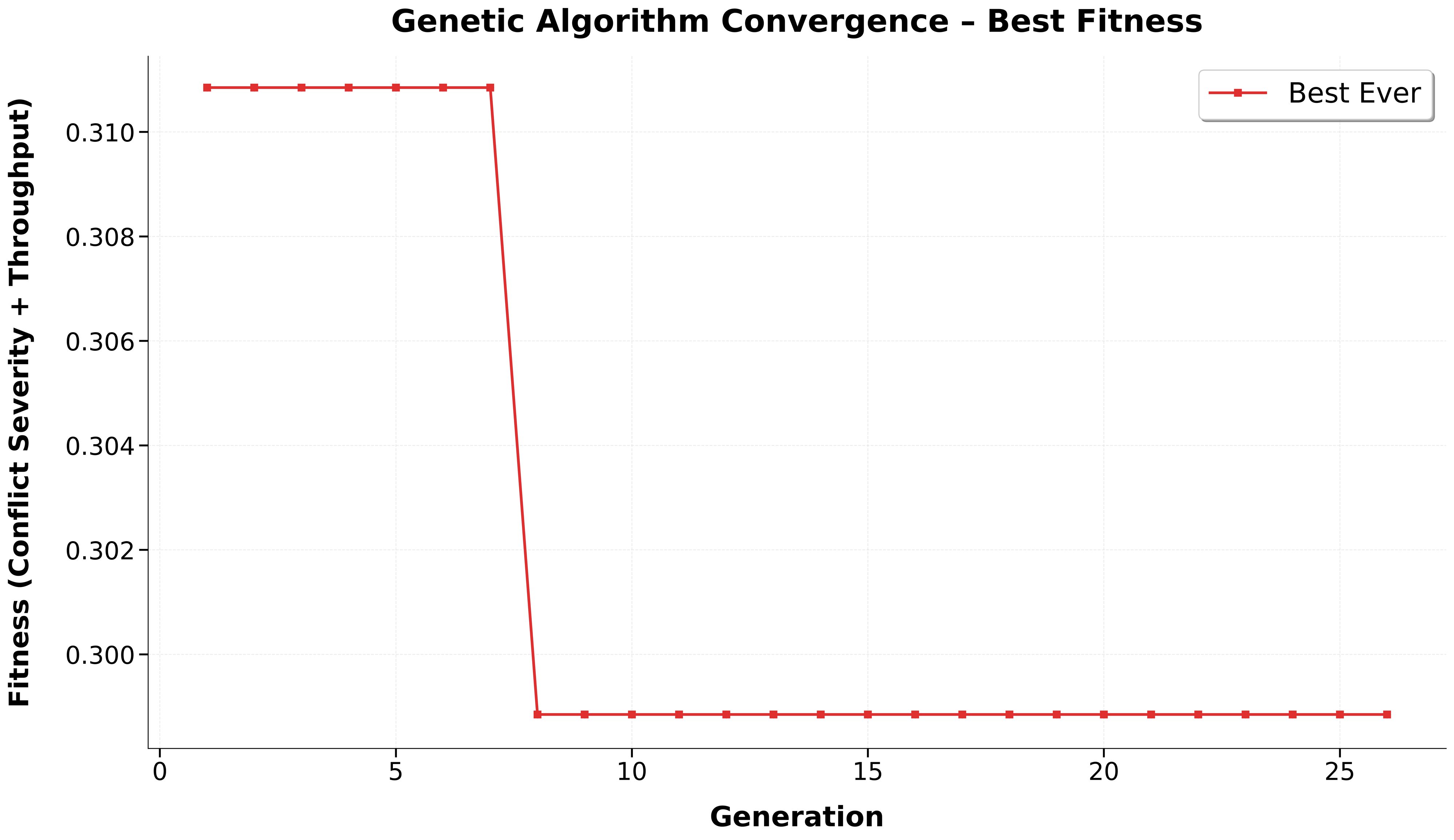}
    \caption{Convergence of the genetic algorithm objective value over successive generations.}
    \label{fig:ga_convergence}
\end{figure}

The genetic algorithm results demonstrate that high-performing street layouts can be identified without evaluating the full design space. Figure~\ref{fig:ga_convergence} shows the convergence of the objective value across generations. The algorithm rapidly improves the population and reaches the Pareto region after evaluating only a fraction of the feasible designs.

In this case study, the genetic algorithm evaluated 43 out of approximately 200 feasible designs before identifying a high-performing layout. The resulting design includes 6-ft dedicated bike lanes, 4-ft sidewalks, 14-ft vehicle lanes, and a balanced directional split with two eastbound lanes and one westbound lane. This design lies near the Pareto frontier and provides a strong balance between safety and throughput for the demand scenario considered.

\subsubsection{Interpretation of optimal designs}
The optimized street layouts reveal several important design insights. First, the highest-performing designs do not simply maximize vehicle capacity. Instead, they allocate the available right-of-way to balance operational efficiency with safety, reducing severe multimodal conflicts while maintaining adequate vehicle throughput. Second, dedicated bicycle facilities substantially reduce high-severity interactions between bicycles and motor vehicles, highlighting their importance in complete street design. Third, the allocation of directional vehicle lanes has a significant influence on both traffic operations and safety by improving the ability to accommodate directional demand while reducing disruptive lane-changing and passing maneuvers.

The results also underscore the importance of evaluating street designs under realistic stochastic traffic conditions. A layout that performs well under average demand may become substantially less effective when curbside stopping, pedestrian crossings, jaywalking, and micromobility interactions occur simultaneously. By explicitly representing these stochastic events within the microscopic simulation, the proposed framework identifies street designs that are not only efficient under typical conditions but also more robust to the day-to-day variability of urban traffic.

\section{Conclusions}
This paper introduced a multi-agent microscopic simulation framework that transforms Complete Street design from qualitative guidelines into a quantitative methodology. By integrating conventional traffic-flow models with detailed urban interaction models, the simulator captures the behaviors of heterogeneous road users, including motorized vehicles, transit, micromobility users, and pedestrians sharing a constrained right-of-way. Operating within a Monte Carlo framework to account for the stochastic nature of urban traffic, the simulator evaluates street designs along two competing dimensions: vehicle throughput and a physics-based conflict severity metric that explicitly accounts for differences in road-user vulnerability.

Application of the framework to a high-activity corridor on Belmont Avenue in Chicago, Illinois, demonstrated its ability to capture the nonlinear operational and safety impacts of modern urban friction, including double parking, jaywalking, and mixed-autonomy traffic. The numerical results highlights the trade-offs between safety and operational efficiency in street design. By coupling the simulator with a genetic algorithm, the framework efficiently explored a complex multimodal design space and identified high-performing, Pareto-efficient street layouts with substantially fewer evaluations than exhaustive enumeration.

Ultimately, this study demonstrates that multimodal street design can be formulated as a simulation-based multi-objective optimization problem. By linking microscopic road-user interactions to macroscopic corridor performance measures, the proposed framework provides public agencies and urban planners with a predictive virtual testbed for evaluating alternative street designs, balancing competing policy objectives, and supporting infrastructure investment decisions before physical implementation. Future work will extend the framework to signalized urban networks, incorporate video-based computer vision for behavioral calibration, and investigate adaptive curbside management strategies for real-time street operation.

\section*{ACKNOWLEDGMENTS}

This research was supported by the Center for Connected and Automated Transportation (CCAT), a regional University Transportation Center funded by the United States Department of Transportation (USDOT).
The authors wish to express their deepest gratitude to our co-author, mentor, and colleague, Professor Hani S. Mahmassani, who passed away during the preparation of this work. His pioneering vision for intelligent transportation systems, relentless pursuit of scientific excellence, and warm mentorship profoundly shaped this research and continue to inspire everyone who had the privilege of working with him. This paper is dedicated to his memory and enduring legacy in transportation science.

\bibliographystyle{trb}
\bibliography{trb_template}

\end{document}